\documentstyle[psfig,preprint,aps]{revtex}
\begin{document}
\draft
\preprint{MKPH-T-97-4}
\title{
{\bf Coherent $\eta$-photoproduction on $^4$He and $^{12}$C 
in the near-threshold region}
\footnote{Supported by the Deutsche Forschungsgemeinschaft (SFB 201)}}

\author{A.\ Fix\footnote
{Supported by the Deutscher Akademischer Austauschdienst} 
and H.\ Arenh\"ovel}
\address{
Institut f\"{u}r Kernphysik, Johannes Gutenberg-Universit\"{a}t,
       D-55099 Mainz, Germany}
\date{\today}
\maketitle

\begin{abstract}

Coherent $\eta$ meson photoproduction on $^4$He and $^{12}$C 
is considered in the near-threshold region. The elementary
$\eta$ photoproduction operator includes contributions from the  
$S_{11}(1535)$ and $D_{13}(1520)$ resonances as well as 
$t$-channel vector meson exchange and the nucleon pole terms. Due to the 
suppression of the dominant $S_{11}(1535)$ resonance for spin and isospin 
saturated nuclei, the reaction is mainly governed by $\omega$ exchange. 
Furthermore, the influence of Fermi motion and of different prescriptions 
for the choice of the invariant reaction energy $W_{\gamma N}$ in the 
elementary amplitude is studied. 
\end{abstract}
\pacs{PACS number(s): 13.60 Le, 25.20 Lj}

\section{Introduction}
During the past few years a large number of experimental and theoretical
investigations of $\eta$ meson photoproduction on nucleon and nuclei 
has been undertaken. Because of the isoscalar nature of the $\eta$ meson, 
these reactions yield unique information concerning the properties of 
baryon resonances with isospin T=1/2. Most of the past theoretical work 
was devoted to the study of the $S_{11}(1535)$ resonance and its 
properties in nuclear matter \cite{Kru95,Mukh95,Sau95,Breit97,Car93}. 
This resonance has a dominant 
contribution in $\eta$ production on the nucleon near 
threshold because of its strong coupling to the $\eta N$ channel. As a 
consequence, the observation of other resonances and of the nonresonant terms 
is possible only via their interference
with the dominant $S_{11}(1535)$ excitation amplitude. Therefore,
theoretical predictions about their properties extracted from the 
$\gamma N \rightarrow \eta N$ analysis depend considerably on
the $S_{11}(1535)$ parameters which are not well known yet.
This problem may be solved by studying $\eta$ photoproduction on nuclei 
which can be used as spin-isospin filters allowing one 
to investigate the individual parts of the elementary amplitude.

In this spirit, the coherent $\eta$ photoproduction on spin-isospin 
saturated nuclei is of special interest, because in these processes the 
$S_{11}(1535)$ resonance is suppressed, and thus the corresponding 
cross sections are sensitive to the ``small'' parts of the elementary 
photoproduction operator. 
On the other hand, a disadvantage of such reactions is the difficulty 
of their experimental isolation due to the smallness of the coherent cross 
section compared to the background quasifree process 
$A(\gamma,\eta )$. As a consequence, rather little effort has been 
devoted to their theoretical 
study. This situation will probably change in the near future
with new measurements of $\eta$ photoproduction on nuclei 
with c.w.\ electron machines, like for example, on $^4$He with MAMI 
\cite{Str}. In anticipation of such experiments, we present here 
theoretical results for $\eta$ photoproduction on spin-isospin
saturated nuclei, for which we have chosen as specific examples 
$^4$He and $^{12}$C. 

An early, rather detailed study of these processes was performed by 
Bennhold and Tanabe \cite{Bennh91}. Their elementary model, based 
on a coupled channel approach, gave a satisfactory description of the at 
that time available data on the $\gamma p \rightarrow \eta p$ reaction, 
which, however, were not very precise.
Their approach is a pure isobar model and does not include any 
background terms. On the other hand, such terms {\it a priori} 
give nonnegligible contributions to the elementary amplitude and 
their role in coherent reactions  may be important. Therefore, with the
appearance of new precise $\gamma p \rightarrow 
 \eta p$ data \cite{Kru95} it is interesting to review the theoretical 
results for $\eta$ photoproduction on such nuclei by including such 
background terms.

Our treatment of the elementary process follows
the effective Lagrangian approach used for the analysis
of meson photoproduction by many authors \cite{Mukh95,Breit97,Bennh89}. 
The $\gamma  N \rightarrow \eta N$ amplitude 
is obtained by means of second-order Feynman diagrams for the nucleon
pole and vector mesons exchange as well as for the resonant states
which are expected to give significant contributions to 
$\eta$ photoproduction
in the near-threshold region. The advantage of this approach is that it 
gives a production operator valid for an arbitrary frame in a rather 
simple analytical form which is convenient
for the implementation into the nuclear $(\gamma,\eta)$ process.
The ingredients of the model are presented in Sect.\ 2. In Sect.\ 3
we discuss our results for the coherent reactions 
on $^4$He and $^{12}$C, and compare them with those obtained by other 
authors. Conclusions are given in Sect.\ 4.

\section{The elementary process}
The details of the kinematics and the structure of the $\eta$ photoproduction 
amplitude on the nucleon have been described in 
\cite{Mukh95,Breit97,Bennh91}. Therefore, we give in this section 
only a brief summary of the elementary reaction 
\begin{equation}\label{g1}
\gamma(k_{\mu},\vec{\varepsilon}_\lambda)+N(p_{\mu})\rightarrow\eta(q_{\mu})
+N(p_{\mu}^{ \prime})\,,
\end{equation}
in order to establish the notation and to describe the ingredients.
Throughout the paper, the conventions of Bjorken and Drell \cite{Bjor64} 
are used. The four-momenta of the participating particles are denoted by 
\begin{equation}
k_{\mu}=(k_0,\vec{k}\,),\quad  q_{\mu}=(q_0,\vec{q}\,),\quad  
p_{\mu}=(p_0,\vec{p}\,),\quad p_{\mu}^{\prime}=(p_0^{\prime},
\vec{p}^{\,\prime}) \,,
\end{equation}
for photon, $\eta$ meson, initial and final nucleon, respectively, and the 
photon polarization vector by $\vec{\varepsilon}_\lambda$. The particle 
energies are 
\begin{equation}
k_0=k, \quad q_0=\sqrt{q^2+m_{\eta}^2}, \quad p_0^{(\prime)}
=\sqrt{p^{(\prime)2}+m^2}\,,
\end{equation}
where $m$  and $m_{\eta}$  stand for the masses of nucleon and $\eta$ meson, 
and $k$, $q$, $p$, and $p^{\prime}$ denote the absolute values of the 
respective three-momenta.

The $S$-matrix of the process (\ref{g1}) is given by
\begin{equation}
S_{fi}= \delta_{fi}-i(2\pi)^4
\delta^{ 4}(p_{\mu}^{\prime}+
q_{\mu}-p_{\mu}-k_{\mu})
\sqrt{\frac{m^2}{2k_02q_0p_0p^{ \prime}_0}} T_{fi}\,,
\end{equation}
where the transition matrix element $T_{fi}$ can be written 
in terms of CGLN amplitudes \cite{CGLN}
\begin{equation}\label{g2}
T_{fi}= T_{s's\lambda}=\bar{u}_f(p^{\prime},s'\,)\Big[\sum^4_{j=1}
A_j(s,t,u)M_{j\lambda}\Big]u_i(p,s).
\end{equation}
Here, $u_{i}(p,s)$ and $u_{f}(p',s')$ denote the covariantly normalized 
Dirac spinors of the intial and final nucleon, respectively, $M_{j\lambda}$ 
the gauge invariant CGLN operators and $A_j$ the invariant scalar amplitudes
which depend on the Mandelstam variables
\begin{eqnarray}
          s&=&(k_{\mu}+p_{\mu})^2\,,\quad
          t=(k_{\mu}-q_{\mu})^2\,,\quad
          u=(k_{\mu}-p_{\mu}^{ \prime})^2\,. 
\end{eqnarray}
The polarization averaged $\gamma N \rightarrow \eta N$
differential cross section in the $\gamma N$ c.m.\ system is related to
the $T$-matrix (\ref{g2}) by
\begin{equation}
\frac{d\sigma}{d\Omega^{c.m.}}=\frac{1}{(4\pi)^2} \frac{q}{k}
 \frac{m^2}{W_{\gamma N}^2}\frac{1}{4}\sum_{s's\lambda}|T_{s's\lambda}|^2,
\end{equation}
where $W_{\gamma N}=\sqrt{s}$ is the $\gamma N$ invariant mass.

\begin{figure}
\centerline{\psfig{figure=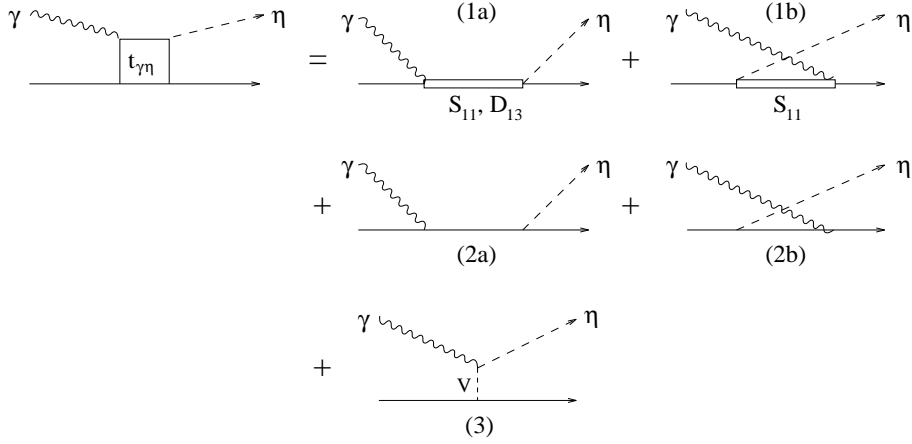,width=12cm,angle=0}}
\vspace*{.5cm}
\caption{Diagrams of all contributions to the elementary amplitude: 
$S_{11}$ and $D_{13}$ resonances in $s$- (1a) and $S_{11}$  resonance in $u$-channel (1b), nucleon pole terms in $s$- (2a) and $u$-channel (2b), 
and $t$-channel vector meson (V) exchange (3). 
}
\label{fig1}
\end{figure}

Fig.\ \ref{fig1} shows the Feynman diagrams included in our calculation. The
corresponding vertex functions and propagators are listed in Tables 
\ref{tab1} and \ref{tab2}. For the $\eta NN$ vertex we use pseudoscalar 
coupling with $g_{\eta NN}^2/4\pi$=0.4 in agreement with \cite{Tiat94,KiT96} 
in the analysis of experimental
$\gamma p \rightarrow \eta p$ angular distributions \cite{Kru95}. 
All parameters for the vector meson $t$-channel exchange 
with $\rho$ and $\omega$ mesons are given in Table \ref{tab3}. We use 
the empirical hadronic coupling constants from the compilation of 
Dumbrajs {\it et al}. \cite{Dum83}. 
The radiative couplings $\lambda_V$ $(V=\rho,\omega)$ 
may be extracted from the electromagnetic 
decay widths $\Gamma(V\rightarrow\gamma\eta)$ \cite{Dol,PDG96} and 
were determined 
in Ref.\ \cite{Mukh95}. 
A monopole form factor, 
\begin{equation}
F(t)=\frac{\Lambda_V^2-m_V^2}{\Lambda_V^2-t}\,,
\end{equation}
with $\Lambda_V^2=2m_V^2$ has been introduced at the $VNN$ vertices. 

The resonance part of the amplitude is represented
by two $T = 1/2$ resonances - $S_{11}(1535)$ and $D_{13}(1520)$ - 
which contribute significantly to $\eta$ photoproduction in the 
near-threshold
region. The other resonances can be neglegted because of their large 
masses and/or small couplings to the $\eta N$ channel \cite{Mukh95}. 
Also the Roper resonance $P_{11}(1440)$ is omitted in our calculation 
because it is below the $\eta N$ threshold and 
its coupling to this channel is not well determined. 
The analysis of the new Mainz data \cite{Kru95} indicates anyway 
a small role of this resonance in $\eta$ production \cite{Mukh95}. 
The relevant resonance parameters are listed 
in Table \ref{tab3}. The mass and width as well as the helicity couplings 
for the $D_{13}(1520)$ were taken from the 1996 PDG listings \cite{PDG96}. 
The corresponding values for the dominant $S_{11}(1535)$ resonance 
were considered as adjustible parameters
in fitting the new data on the total $\gamma p \rightarrow \eta p$ cross 
section \cite{Kru95}. The ratio $A_{1/2}^n/A_{1/2}^p$ of the neuteron 
to the proton amplitude of the $S_{11}(1535)$ photoexcitation 
was taken from Ref.\ \cite{Fix97} as obtained from the analysis of the 
inclusive $d(\gamma,\eta)$ cross section, where the rescattering
effects in the final state were taken into account. Furthermore, for 
$S_{11}(1535)$ we use an energy dependent decay width 
\begin{equation}\label{2a}
\Gamma(W_{\gamma N})=\Gamma(M_{S_{11}})\Big(b_{\eta}\frac{q}{q^{*}}
+b_{\pi}\frac{q_{\pi}} {q_{\pi}^{*}}+b_{\pi\pi}\Big),
\end{equation}
with the branching ratios $b_{\eta}=0.5$, $b_{\pi}=0.4$ and $b_{\pi\pi}=0.1$. 
Here $q$ and $q_{\pi}$ are the $\eta$ and $\pi$ meson momenta in the 
$\gamma N$ c.m.\ system as functions of $W_{\gamma N}$, 
and $q^{*}$ and $q_{\pi}^{*}$ 
are the respective momenta at $W_{\gamma N}=M_{S_{11}}$. 

With respect to the $D_{13}(1520)$ resonance, it is known that the 
relativistic treatment of a $J=3/2$ particle contains an ambiguity related 
to its off-mass-shell extrapolation \cite{Will85}. As a consequence, the 
implementation of the free Rarita-Schwinger propagator in the intermediate 
off-mass-shell state leads to the appearance of a resonance contribution
in the nonresonant amplitudes. In order to avoid this ambiguity,
following Ref.\ \cite{Adel85}, we have replaced the resonance mass 
$M_{D_{13}}$ in the numerator of the $D_{13}(1520)$ propagator by the 
invariant energy $W_{\gamma N}$ (see Table \ref{tab2}). Furthermore, 
we have omitted the $D_{13}(1520)$ formation in the $u$ channel, 
where the resonance is always far off-shell. We will also 
consider medium modifications of the width when implementing the elementary 
amplitude in a nucleus as will be discussed in Sect.\ 3. 

The invariant amplitudes $A_i$, corresponding to the individual terms
in Fig.\ \ref{fig1}, are collected in the Appendix. In Fig.\ \ref{fig2} 
the predictions of our elementary model for the $\gamma p \rightarrow
 \eta p$ differential cross section are compared with new Mainz data 
\cite{Kru95}. One readily notes that a satisfactory agreement is achieved. 
We would like to piont out that within our parametrization 
the contributions from nucleon pole and vector meson exchange cancel
each other almost completely, leaving practically the whole angular
distribution to be given by the $S_{11}(1535)$ and $D_{13}(1520)$ resonances. 

For the calculation of $\eta$ photoproduction on a nucleus 
with nonrelativistic wave functions, it is necessary to express the 
photoproduction amplitude in terms of Pauli matrices and two-component 
spinors. Therefore, we introduce a corresponding $t$-operator by
\begin{equation}\label{2g}
\chi^{\dagger}_{s'} t_{\lambda} \chi_s =
\bar{u}_f(p',s'\,)\Big[\sum_{j}A_jM_{j\lambda}\Big]u_i(p,s)\,,
\end{equation}
and decompose it into a so-called non-spin-flip and a spin-flip part 
according to 
\begin{equation}\label{1a}
t_{\lambda}=K(\vec{\varepsilon}_{\lambda})
+i\vec L(\vec{\varepsilon}_{\lambda})\cdot\vec{\sigma}.
\end{equation}
The amplitudes $K$ and $\vec{L}$ can be written in the following form
\begin{eqnarray}\label{K}
K(\vec{p}^{\,\prime},\vec{p},\vec{k},\vec{\varepsilon}) &=& 
F_1\Big(\frac{\vec{p}\cdot(\vec{\varepsilon}\times\vec{k})}{p_0+m}-
\frac{\vec{p}^{\,\prime}\cdot(\vec{\varepsilon}\times\vec{k})}
{p_0^{ \prime}+m}\Big) \nonumber \\ 
 &+& F_2\frac{\vec{p}^{\,\prime}\cdot(\vec{\varepsilon}\times\vec{p})}
{(p_0+m)(p_0^{ \prime}+m)}
+F_3\frac{\vec{p}^{\,\prime}\cdot(\vec{k}\times\vec{p})}
{(p_0+m)(p_0^{ \prime}+m)}\, , 
\end{eqnarray}
\begin{eqnarray}\label{L}
\vec{L}(\vec{p}^{\,\prime},\vec{p},\vec{k},\vec{\varepsilon}) &=&
\vec{\varepsilon} \Big[F_1\Big(2k_0-\frac{\vec{k}\cdot\vec{p}}
{p_0+m}-\frac{\vec{k}\cdot\vec{p}^{\,\prime}}{p_0^{ \prime}+m}\Big)+
F_2\Big(\frac{\vec{p}\cdot\vec{p}^{\,\prime}}
{(p_0+m)(p_0^{ \prime}+m)}-1\Big)\Big] \nonumber \\ 
&+&
\vec{k} \Big[F_1\Big(\frac{\vec{\varepsilon}\cdot\vec{p}}
{p_0+m}+\frac{\vec{\varepsilon}\cdot\vec{p}^{\,\prime}}
{p_0^{ \prime}+m}\Big)+
F_3\Big(\frac{\vec{p}\cdot\vec{p}^{\,\prime}}
{(p_0+m)(p_0^{ \prime}+m)}-1\Big)\Big] \nonumber \\ 
 &+&
\vec{p} \Big[\frac{F_4}{p_0+m}-
F_2\frac{\vec{\varepsilon}\cdot\vec{p}^{\,\prime}}{(p_0+m)(p_0^{ \prime}+m)}-
F_3\frac{\vec{k}\cdot\vec{p}^{\,\prime}}{(p_0+m)
(p_0^{ \prime}+m)}\Big] \nonumber \\ 
 &+&
\vec{p}^{\,\prime} \Big[\frac{F_4+2k_0F_3}{p_0^{ \prime}+m}-
F_2\frac{\vec{\varepsilon}\cdot\vec{p}}{(p_0+m)(p_0^{ \prime}+m)}-
F_3\frac{\vec{k}\cdot\vec{p}}{(p_0+m)(p_0^{ \prime}+m)}\Big]
\end{eqnarray}
where $\vec{p}^{\,\prime}=\vec{p}+\vec{k}-\vec{q}$, and $\vec{\varepsilon}$ 
denotes the photon polarization vector. Furthermore, the 
$F_i$ are expressed in terms of $A_i$ as follows
\begin{eqnarray}
F_1 &=& N^{ \prime}N (A_1-2mA_4)\,,  \\
F_2 &=& N^{ \prime}N \Big[k_0(A_1-2mA_4)+
A_3k^{\mu} (p_{\mu}^{ \prime}-p_{\mu})-
A_4k^{\mu} (p_{\mu}^{ \prime}+p_{\mu})\Big]\,, \\
F_3 &=& N^{ \prime}N \Big[ A_3
\vec{\varepsilon}\cdot(\vec{p}^{\,\prime}-\vec{p})-A_4
\vec{\varepsilon}\cdot(\vec{p}^{\,\prime}+\vec{p})\Big]\,, \\
F_4 &=& N^{ \prime}N \Big[ 2A_2
(\vec{\varepsilon}\cdot\vec{p}^{\,\prime} k^{\mu}p_{\mu}-
\vec{\varepsilon}\cdot\vec{p} k^{\mu}p_{\mu}^{ \prime})-
k_0A_3\vec{\varepsilon}\cdot(\vec{p}-\vec{p}^{\,\prime})-
k_0A_4\vec{\varepsilon}\cdot(\vec{p}+\vec{p}^{\,\prime})\Big]\,,
\end{eqnarray}
with $N = \sqrt{(p_0+m)/2m}$,
$N^{ \prime} = \sqrt{(p_0^{ \prime}+m)/2m}$.

When applying the elementary operator (\ref{1a})-(\ref{L})
to a bound nucleon, which is always off-mass-shell, 
the energies $p_0$ and $p_0^{\prime}$ are not defined in a 
nonrelativistic framework, and thus the problem of assigning an energy 
to a bound nucleon arises. This question will be addressed in the 
next section.     

The spin-dependent amplitude $\vec{L}$, which is dominated by the
$S_{11}(1535)$ excitation, plays a major role in 
$\eta$ photoproduction on a free nucleon. However, due to the
spin-isospin selection rules, the coherent $(\gamma,\eta)$ process 
on $J = T = 0$ nuclei is essentially determined by the isoscalar 
part of the non-spin-flip amplitude $K^{(0)}$. Therefore, the influence
of the $S_{11}(1535)$ resonance is largely suppressed in this case. 
It should be noted, however, that by the spin decomposition of 
the elementary operator in 
the $\eta N$ c.m.\ system, the contribution from $S_{11}(1535)$, which
determines the electric dipole amplitude $E_{0^{+}}$, enters only into the 
spin-dependent component $\vec{L}$. But in general for an 
arbitrary frame, a small contribution from this resonance is 
mixed into the spin-independent part $K$. Although this mixing normally is 
insignificant relative to the $\vec{L}$-term, the resulting effect may be
noticeable if the spin-flip term is suppressed, 
because of the relatively large strength of the $S_{11}(1535)$ 
contribution in the $\gamma N \rightarrow \eta N$ amplitude.  

\section{Coherent $\eta$ photoproduction on nuclei}

Now we will consider the near-threshold coherent $\eta$ photoproduction 
on a nucleus 
\begin{equation}\label{4f}
\gamma + A \rightarrow \eta +A 
\end{equation}
with mass number $A$, spin $J$=0 and isospin $T$=0. In this case, the 
differential cross section in the $\gamma$-nucleus c.m.\ frame can be 
written as
\begin{equation}\label{5f}
\frac{d\sigma}{d\Omega^{c.m.}}=\frac{1}{(4\pi)^2}\, \frac{q}{k}\,
 \frac{E_{A,k}E_{A,q}}{W_{\gamma A}^2}\, \frac{1}{2} 
\sum_{\lambda}|T_{A,\lambda}^{c.m.}(\vec k,\,\vec q\,)|^2,
\end{equation}
where $W_{\gamma A}$ denotes the invariant mass of the $\gamma A$ system 
and $E_{A,k}$ and $E_{A,q}$ the initial and final nuclear energies, 
respectively, i.e., $E_{A,p}=\sqrt{p^2+M_A^2}$ with the nuclear mass 
$M_A$. The nuclear states are noncovariantly normalized to unity. In the 
laboratory frame the cross section is given by 
\begin{equation}\label{5fl}
\frac{d\sigma}{d\Omega^{lab}}=\frac{1}{(4\pi)^2}\, \frac{q_L^2}{k_L}\,
 \frac{E_{A,q_L}}{q_L(M_A+k_L)-k_LE_{q_L}\cos\theta_L}\, \frac{1}{2}
\sum_{\lambda}|T_{A,\lambda}^{lab}(\vec k_L,\,\vec q_L\,)|^2,
\end{equation}
where all quantities refer to the laboratory frame. In order to check the 
frame independence of the calculation, we will compare the direct 
evaluation of (\ref{5f}) with the calculation obtained in the lab frame from 
(\ref{5fl}) and subsequently transformed with the Jacobian
\begin{equation}\label{jac}
\frac{d\Omega^{lab}}{d\Omega^{c.m.}}=\frac{q^2}{q_L}\,
\frac{M_A}{E_{A,k}q+E_q k \cos\theta}
\end{equation}
to the c.m.\ frame. 

In the impulse approximation (IA), the nuclear photoproduction 
amplitude is given as a sum over the elementary amplitudes of all 
nucleons in the nucleus. For example, in the c.m.\ frame, the transition matrix 
$T^{c.m.}_{A,\lambda}$ can be represented in the form
\begin{equation}\label{6f}
T_{A,\lambda}^{c.m.}(\vec k,\,\vec q\,)=
A\int d^3p\,d^3q^{ \prime}\,\chi_{\vec{q}}^{(-)*}(\vec{q}^{\,\prime})
K^{(0)}(\vec{p}-\frac{\vec{k}}{A}+\vec{Q}^{\,\prime},
\vec{p}-\frac{\vec{k}}{A},\vec{k},
\vec{\varepsilon}_{\lambda})
\rho_A(\vec{p} + \vec{Q}^{\,\prime},\vec{p}\,) \,,\\
\end{equation}
where $\vec{Q}^{\,\prime}=\vec{k} - \vec{q}^{\,\prime}$.
Here $\rho_A(\vec{p}^{\,\prime},\vec{p})$ denotes the one-body nuclear 
density matrix of the nuclear ground state in momentum space 
\begin{equation}
\rho_A(\vec{p}^{\,\prime},\vec{p})=\langle A|a^\dagger(\vec p^{\,\prime})
a(\vec p\,)|A\rangle \,,\label{rho1}
\end{equation}
where $a^\dagger(\vec{p}\,)$ and $a(\vec{p}\,)$ denote creation and 
anihilation operators for a bound nucleon with momentum $\vec{p}$ in the 
nuclear rest frame. 
The isoscalar part $K^{(0)}$ of the spin-independent amplitude 
(\ref{K}) is defined as $K^{(0)}=1/2(K_p+K_n)$ with index $p$ referring
to the proton and $n$ to the neutron. The function 
$\chi_{\vec{q}}^{(-)}(\vec{q}^{\,\prime})$ describes the relative 
motion of the outgoing $\eta$ with respect to the recoiling nucleus 
including a final state interaction via an optical potential. 
Since we are interested in the main 
properties of the reaction (\ref{4f}), we simplify further and ignore 
here the $\eta$-nucleus interaction. This means, we substitute 
$\chi_{\vec{q}}^{(-)}(\vec{q}^{\,\prime})$ by the free meson function 
$\delta(\vec{q}-\vec{q}^{\,\prime})$
in the matrix element (\ref{6f}) leading with (\ref{rho1}) to 
\begin{equation}\label{7f}
T_{A,\lambda}^{c.m.}(\vec k,\,\vec q\,)=
A \int d^3p\, K^{(0)}
(\vec{p}-\frac{\vec{k}}{A}+\vec{Q},
\vec{p}-\frac{\vec{k}}{A},\vec{k},
\vec{\varepsilon}_{\lambda})\rho_A(\vec{p} + \vec{Q},\vec{p})\,,
\end{equation}
where $\vec{Q}=\vec{k} - \vec{q}$ is the transferred momentum. 

In order to avoid the complication of the nuclear Fermi motion in the 
evaluation of (\ref{7f}), a further approximation is 
widely used in the study of meson photoproduction 
on light nuclei \cite{Bennh91,Kam87} which allows one to express the 
matrix element in terms of the nuclear body form factor $F_A(Q)$. It 
is the so-called factorization approximation \cite{Bennh91} which we give 
here with respect to the laboratory frame for convenience 
\begin{equation}\label{8f}
T^{lab}_{A,\lambda}(\vec k_L,\,\vec q_L\,)=A F_A(Q_L) 
K^{(0)}(\vec{p}_{eff}+\vec Q_L,\vec{p}_{eff}, \vec{k_L},
\vec{\varepsilon}_{\lambda})\,,
\end{equation}
where the elementary operator is frozen at the average effective nucleon 
momentum in the nuclear rest system, i.e., the lab system
\begin{equation}\label{Peff}
                  \vec{p}_{eff}=-\frac{A-1}{2A}\vec{Q}_L\,,
\end{equation}
which corresponds to the following assignments in the $\gamma A$ c.m. frame 
\cite{Kam87}
\begin{eqnarray}
\vec p &=& -\frac{1}{A}\vec k -\frac{A-1}{2A}\vec{Q}\,,\\
\vec p^{\,\prime} &=& -\frac{1}{A}\vec q + \frac{A-1}{2A}\vec{Q}\,.
\end{eqnarray}
In this approximation, the nucleon energies $p_0$ and $p_0'$ are taken on 
shell. The exact treatment of the Fermi motion by a direct 
computation of the three-dimensional integral in the matrix element of 
(\ref{7f}) will be discussed below for $^4$He. 

For $^{12}$C we have taken a phenomenological fit for the charge form factor 
$F_{^{12}C}^{ch}(Q)$ from \cite{Pol73}
\begin{equation}\label{10f}
     F_{^{12}C}^{ch}(Q)=-\frac{3\pi b\:[\cos(QR)-(\pi b/R)\sin(QR)
           \coth(\pi bQ)]}{QR^2 \sinh(\pi bQ)[1+\pi^2b^2/R^2]} ,
\end{equation}
with two parameters $R=2.179$ fm and $b=0.503$ fm. The body form factor
is then obtained by
\begin{equation}
F_{^{12}C}(Q)=[F_p(Q)]^{-1}F_{^{12}C}^{ch}(Q)\,,
\end{equation}
with the proton form factor $F_p(Q)$ taken in the dipole parametrization
\begin{equation}
       F_p(Q)=\frac{1}{(1+0.055\,(Q/\mbox{fm}^{-1})^2)^2}.
\end{equation}

For $^4$He we use as wave function a product ansatz which consists of a 
relative motion of the active nucleon with respect to the spectator 
nucleons building a $3N$ cluster with an internal wave function. 
The relative motion is described by an $s$-wave $\psi(\vec{r})$ with 
$\vec{r}$ being the relative coordinate in the $N-(3N)$ system. The 
internal cluster wave function is not needed since we neglect 
antisymmetrization between the active nucleon and the $3N$-cluster. 
For $\psi(\vec{r})$ we have taken from \cite{Sher83} 
the phenomenological form 
\begin{equation} \label{fHe} 
\psi(r)=N\frac{e^{-\alpha r}}{r} \sum^5_{j=1}a_je^{-\beta_jr}\ , 
\end{equation}
where $\alpha = 0.846\,\mbox{fm}^{-1}$ and the parameters $a_j$ 
and $\beta_j$ are listed in Table \ref{tab4}.
The corresponding body form factor, which can be written as
\begin{equation} 
F_{^4He}(Q)=\int d^3r|\psi(r)|^2 e^{i\frac{3}{4}\vec{Q}\cdot\vec{r}}=
4\pi N^2 \frac{4}{3Q}\sum_{ij}a_ia_j \arctan\frac{3Q}{4(2\alpha+
\beta_i+\beta_j)},
\end{equation}   
fits the measured $^4$He charge form factor up to its second maximum.
The functional form (\ref{fHe}) allows also a simple analytic expression 
for the momentum space representation 
\begin{equation} 
\psi(p)=\int d^3r \psi(r) e^{i\frac{3}{4}\vec{p}\cdot\vec{r}}=
4\pi N \sum_{j}\frac{a_j}{(\alpha+\beta_j)^2+(3p/4)^2}\,,
\end{equation}
which is needed for the explicit evaluation of the Fermi motion in the 
integral of (\ref{7f}). 

We begin the discussion of our results with the total cross sections for 
the coherent reactions $^4$He$(\gamma,\eta)^4$He and 
$^{12}$C$(\gamma,\eta)^{12}$C which are shown in Fig.\ \ref{figTot}.
One readily notes that the coherent cross section on $J=T=0$ nuclei is 
very small in magnitude, of the relative order of $10^{-3}$, compared to the 
incoherent process (see for example \cite{KruPR95}). 
This is mainly due to the suppression of the in the 
elementary reaction dominant $S_{11}(1535)$ resonance. 
Another reason is that the coherent nuclear transition amplitude via 
the body form factor falls off rapidly at the high momentum 
transfers associated with the $\eta$ production. Although the coherence 
factor $A^2$ is nine times bigger for $^{12}$C compared to $^4$He, the 
cross section for $^{12}$C is considerably smaller because of the much 
faster fall-off of the body form factor with momentum transfer $Q$. 

The contributions of the individual terms of our elementary 
$\gamma N \rightarrow \eta N$ model are illustrated in Fig.\ \ref{C12_1} 
for the differential cross section of $^{12}$C. 
As mentioned before, the $S_{11}(1535)$ resonance has some influence on the 
non-spin-flip part of the elementary operator. But as one can see on the 
right hand panel of Fig.\ \ref{C12_1}, this effect remains
negligible, since the isoscalar part of the $S_{11}(1535)$ amplitude, 
used in our calculation, is rather small ($T^{(0)}/T_p=0.076$). 
It is obvious that of the vector mesons only the isoscalar 
$\omega$ meson contributes to the amplitude here. In fact, one can see 
on the left hand panel of Fig.\ \ref{C12_1} that this
term dominates the nuclear cross section, while the isoscalar 
part of the nucleon pole terms gives also an insignificant contribution.

As for the $D_{13}(1520)$ resonance, we have taken into account a possible 
modification of its properties in a nuclear environment. 
Experimental results on total photoabsorption \cite{Bian93} 
as well as $\pi^0$ photoproduction on light nuclei \cite{KruPR95} show a 
strong depletion of the resonance structure in the region of the 
$D_{13}(1520)$ resonance. This effect may be attributed to  
a strong shortening of the $D_{13}(1520)$ lifetime in a nucleus, where
additional absorption channels, involving inelastic $NN^{*}$ collisions, 
increase the resonance width. We have simulated such effects in our 
calculations by simply replacing the free resonance width by 
an effective one, i.e., $\Gamma\rightarrow\Gamma^{*}$.
The value $\Gamma^{*}=265$ MeV for the $D_{13}(1520)$ was obtained in 
Ref.\ \cite{Alb94} from the 
analysis of total photoabsorption on $^9$Be and $^{12}$C. 
A questionable point of this treatment 
is that the parameters, extracted from $^9$Be and $^{12}$C data, 
may not be appropriate for such a light nucleus like $^4$He.
However, we believe that the possible error is not dramatic and does not 
lead to a qualitative change of the results. 
This medium modification of the $D_{13}(1520)$ width leads to a noticable 
damping of its contribution to the cross section, as is demonstrated in the 
right hand panel of Fig.\ \ref{C12_1}. In summary, our calculation 
shows a clear dominance of $\omega$ exchange in the coherent cross section for 
spin-isospin saturated nuclei, while the role of the nucleon pole terms 
as well as of the resonances is not important.

Another remark about the role of the $D_{13}(1520)$ resonance is concerned 
with the fact, that $\omega$ exchange contributes to the real part of 
the $\eta$ production amplitude only, while the real part of the amplitude of 
the $D_{13}(1520)$, located near the $\eta$ production threshold, is 
relatively small. For this reason, we find very little interference between 
these amplitudes in the nuclear cross section. This is in marked contrast 
to a very recent result of \cite{PiS97} where a strong constructive 
interference between vector meson exchange and $D_{13}(1520)$ excitation 
increases the cross section of the coherent reaction on $^{40}$Ca 
substantially. The reason is not clear to us. 

Now we turn to the discussion of the exact treatment of the 
nuclear Fermi motion. In the explicit evaluation 
of the integral in (\ref{7f}), one faces the problem of fixing the 
invariant mass of the $\gamma N$ subsystem
\begin{equation}
W_{\gamma N}=\sqrt{(k_0+p_0)^2-(\vec{k}+\vec{p} )^2}.
\end{equation}
In fact, this question is connected with the choice of the energy 
of the active bound nucleon in the off-shell region, where 
$p_0^2\not=m^2+p^2$. The analogous problem for the coherent process 
on the deuteron has already been discussed in Ref.\ \cite{Breit97}. 
There it was shown that various prescriptions for $W_{\gamma N}$ lead 
to significantly different results. It should be noted, however, that 
the $\eta$ photoproduction on the deuteron is dominated by the 
$S_{11}(1535)$ resonance, and thus the corresponding amplitude depends 
directly on $W_{\gamma N}$ appearing in the resonance propagator.
In addition, the choice of $W_{\gamma N}$ has a large influence on the width
$\Gamma_{S_{11}}(W_{\gamma N})$ to which the cross section is 
also very sensitive. 

On the other hand, in view of the dominant $t$-channel $\omega$ exchange
for the coherent $(\gamma,\eta)$ processes considered here, 
one would expect that the cross section will be rather
insensitive to the different assumptions on $W_{\gamma N}$.
In order to investigate this question, we have considered for 
the $^4$He$(\gamma,\eta)^4$He reaction the 
following alternative prescriptions for 
$W_{\gamma N}$ in the $\gamma\,^4$He c.m.\ system:
\begin{eqnarray}\label{11f}
W_{\gamma N}^{(1)} &=& \Big[\frac{1}{E_{A,k}^2}
\Big(\frac{M_A^2}{4}+k_0E_{A,k}-\vec{k}\cdot\vec{p}\Big)^2
-(\vec{k}+\vec{p} )^2\Big]^{1/2}, \\
W_{\gamma N}^{(2)} &=& \Big[(k_0+\sqrt{p^2+m^2})^2-
(\vec{k}+\vec{p} )^2\Big]^{1/2},\label{11g}
\end{eqnarray}
where $E_{A,k}$ is the $^4$He c.m.\ energy in the initial state.
The first choice (\ref{11f}) corresponds to the assumption, 
that all nucleons share equally the total energy of $^4$He in its rest 
system 
\begin{equation} 
                p_0^{lab}=\frac{M_A}{4}.   \nonumber
\end{equation}
In the second expression (\ref{11g}), the active nucleon is taken on-shell 
before the production process. This case has also been considered in 
Ref.\ \cite{Tiat80} for the $^3$He$(\gamma,\pi^{+})^3$H reaction. Our 
results, corresponding to the different choices of $W_{\gamma N}$ are shown 
in Fig.\ \ref{W}. The only term, which can be significantly affected 
by the value of $W_{\gamma N}$ and which can have some noticeable influence 
on the cross section, is the $D_{13}(1520)$ excitation term in our model.
We note that in the whole region of nucleon momentum $\vec{p}$ one has the 
relation 
\begin{equation} 
           W_{\gamma N}^{(1)} < W_{\gamma N}^{(2)} ,  \nonumber
\end{equation}
because the binding energy of the nucleon, which is taken into account 
in (\ref{11f}), decreases its total energy.  
Thus the different choices for $W_{\gamma N}$ lead to different results for 
the magnitude and position of the $D_{13}(1520)$ 
resonance in the nuclear cross section. However, since the resonance 
contribution is small, the influence of this uncertainity in 
$W_{\gamma N}$ is not significant as can be seen in Fig.\ \ref{W}. 

In this figure we also display the $^4$He$(\gamma,\eta)^4$He angular
distribution obtained within the factorization approximation (\ref{8f}),
where the elementary amplitude is taken on-shell. 
The rather small difference between the corresponding 
curves shows, that the exact treatment of Fermi motion differs little 
from the factorization approach which might not be so surprising after all, 
because of the dominance of $\omega$ exchange. Therefore, we use this 
approximation in the following calculations.

The study of $\eta$ photoproduction on the free nucleon
indicates, that the role of $\omega$ exchange in this reaction is 
relatively small, because of the small radiative coupling 
$\lambda_w$ to the $\gamma\eta$-channel.
As a result, the elementary cross section is rather insensitive
to the uncertainities in the $\omega$ parameters. However, from our 
analysis of the coherent $(\gamma,\eta)$ process on $J = T = 0$
nuclei, where the $\omega$ exchange dominates, it is expected 
that these uncertainities will be reflected in the nuclear cross 
section. We therefore have also investigated the sensitivity of the 
$^4$He$(\gamma,\eta)^4$He reaction to the choice of the vector 
coupling constant $g_{\omega NN}^v$ for which one finds a significant 
variation of values in the literature \cite{Dum83}. In our 
calculation we have used $g_{\omega NN}^v=10.3$ which is close to the 
quark model prediction $g_{\omega NN}^v\approx$11.5 \cite{PDG}. 
On the other hand, in Ref.\ \cite{Laget77} a value $g_{\omega NN}^v$=17.5 
has been chosen in order to reproduce the near-threshold 
$\gamma p\rightarrow\pi^0 p$ data. The resulting cross section using 
this value of $g_{\omega NN}^v$ is shown in Fig.\ \ref{W1}. 
As one can see, with the larger coupling constant the cross section increases 
by about a factor of 3, which is roughly the square of the ratio of 
the coupling constants. In principle, this high sensitivity to the 
$\omega$ exchange makes it possible to determine
the value of the $\omega NN$ coupling constant by a precise measurement of
coherent $\eta$ photoproduction on $J = T = 0$ nuclei provided that 
neglected two-body effects are small indeed.

We have furthermore checked the frame independence of the calculation by 
comparing the c.m.\ differential cross section obtained directly in the 
c.m.\ frame with the one evaluated first in the lab frame according to 
(\ref{5fl}) and then transformed with the help of the Jacobian (\ref{jac}) to 
the c.m.\ frame. The resulting cross sections are shown in Fig.\ \ref{figlabcm} 
for $^4$He. Indeed, the differences are very small indicating that c.m.\ 
motion effects are insignificant. We have checked it also 
for $^{12}$C where the differences are completely negligible. 

Finally, we compare in Fig.\ \ref{Bennh} our results 
for $^4$He$(\gamma,\eta)^4$He with those obtained by other authors.
As has been mentioned before, the coherent reactions, considered here, were 
already investigated in Ref.\ \cite{Bennh91}, where the elementary 
amplitude was obtained in the framework of a coupled channel approach. In
this work, the isoscalar $M^{(0)}_{2^-}$ multipole from the $D_{13}(1520)$
resonance plays a dominant role in 
$^4$He$(\gamma,\eta)^4$He reaction, while the nonresonant
terms were omitted. This is quite contrary to our conclusion,
that $\omega$ exchange gives the major mechanism of coherent 
$(\gamma,\eta)$ processes on $J=T=0$ nuclei. Therefore, the cross section 
from \cite{Bennh91} is by a factor 6 larger than our result, as can be seen 
in the left hand panel of Fig.\ \ref{Bennh}. Recently, another evaluation 
of $\eta$ photoproduction on $^4$He in the PWIA approximation has been 
presented in \cite{Tiat94} using the model of \cite{Bennh91}, but with 
inclusion of the background terms considered in the present work. 
In order to compare the results, we have calculated
$^4$He$(\gamma,\eta)^4$He cross section with the same parameters
in the $t$-channel as in Ref.\ \cite{Tiat94}. The corresponding cross 
section, shown in the right hand panel of Fig.\ \ref{Bennh}, 
understimates still the one of \cite{Tiat94} by about 30 percent. 

In order to explain the origin of these differences, we compare in
Fig.\ \ref{Bennh2} the isoscalar part of the CGLN amplitude 
${\cal F}_2$ as predicted in \cite{Bennh91} with our model. This amplitude,  
appearing in the usual form for the photoproduction operator 
\begin{equation}
F=i{\cal F}_1\vec{\sigma}\cdot\vec{\varepsilon}+
{\cal F}_2\vec{\sigma}\cdot\hat{q}
\vec{\sigma}\cdot(\hat{k}\times\vec{\varepsilon})+
i{\cal F}_3\vec{\sigma}\cdot\hat{k}\hat{q}\cdot\vec{\varepsilon}+
i{\cal F}_4\vec{\sigma}\cdot\hat{k}\hat{q}\cdot\vec{\varepsilon},
\end{equation} 
plays a major role in the coherent meson photoproduction on $J=T=0$
nuclei. As one readily notes in Fig.\ \ref{Bennh2}, the value of 
${\cal F}_2^{(0)}$ as given by our calculation is significantly smaller 
than the one of Ref.\ \cite{Bennh91} in the near-threshold
region. As already mentioned, in \cite{Bennh91} the latter is dominated 
by the resonant $M^{(0)}_{2^-}$ multipole which is much larger in comparison
with our one. Furthermore, as was discussed above, we have taken into account
the broadening of the $D_{13}(1520)$ resonance in a nuclear environment, 
reducing further its contribution to $(\gamma,\eta)$ reactions.   

\section{Conclusion}     
We have studied coherent $\eta$ photoproduction 
from the spin-isospin saturated nuclei $^4$He and $^{12}$C. 
The elementary production operator was obtained in the effective
Lagrangian approach.
Two resonances, $S_{11}(1535)$ and $D_{13}(1520)$, as well as nucleon 
pole terms and vector meson exchange in the $t$-channel were included.
One main point of this work was to investigate the influence
of these different contributions on the coherent $(\gamma,\eta)$ reaction 
on nuclei with zero spin and isospin, where the dominant contribution 
from the $S_{11}(1535)$ resonance is suppressed.  
 
Our calculations 
show that the reaction is dominated by the $\omega$ exchange 
in the $t$-channel, while the 
role of the $D_{13}(1520)$ resonance and other terms is rather small. 
Such ``nonresonant'' character 
leads to some peculiarities of the nuclear cross section. 
For example, we see little sensitivity of the theoretical results to the
different assumptions about the invariant mass of the $\gamma N$ subsystem
in the off-shell region. Furthermore, it was found that the influence
of the Fermi motion on the cross section may successfully be 
simulated by the factorization procedure, using an effective nucleon 
momentum for the on-shell elementary amplitude. 

Finally, we would like to emphasize again the dominant role of the 
$\omega$ exchange in the nuclear reaction amplitude. This is in contrast to 
its small contribution to the $\gamma p \rightarrow \eta p$ reaction. 
For this reason, its influence is rather difficult to determine in the 
elementary process, whereas precise experimental data on coherent 
$(\gamma,\eta)$ reactions on $J=T=0$ nuclei may allow a much cleaner study 
of the $\omega$ contribution. 

\acknowledgments
We would like to thank J.\ Friedrich for very useful discussions concerning 
varios parametrizations of nuclear charge form factors. A.\ F.\ thanks R.\ 
Schmidt, M.\ Schwamb and P.\ Wilhelm for fruitful discussions. 

\renewcommand{\theequation}{A.\arabic{equation}}
\setcounter{equation}{0}
\section*{Appendix: Invariant amplitudes
for the reaction $\gamma N \rightarrow \eta N$}
\label{app3}

Using the expressions for propagators and vertex factors given in
Table \ref{tab1}, the various contributions to the invariant amplitudes 
of (\ref{g2}) are as follows: 

i) Nucleon pole
\begin{eqnarray}
A_1&=& ee_Ng_{\eta NN}
\Big(\frac{1}{s-m^2}+\frac{1}{u-m^2}\Big)\,, \\  
A_2&=& 2ee_Ng_{\eta NN}\frac{1}{(s-m^2)(u-m^2)}\,, \\ 
A_3&=&  -eg_{\eta NN}
\frac{\kappa_N}{2m}\Big(\frac{1}{s-m^2}-\frac{1}{u-m^2}\Big)\,, \\ 
A_4&=&  -eg_{\eta NN}
\frac{\kappa_N}{2m}\Big(\frac{1}{s-m^2}+\frac{1}{u-m^2}\Big)\,.
\end{eqnarray}

(ii) Vector meson exchange 
\begin{eqnarray}
A_1&=&  \frac{e\lambda_V}{m_{\eta}}\frac{g_{VNN}^t}{2m}
\frac{t}{t-m_V^2}\,, \\
A_2&=&  -\frac{e\lambda_V}{m_{\eta}}\frac{g_{VNN}^t}{2m}
\frac{1}{t-m_V^2}\,, \\
A_3&=&0\,, \\
A_4&=&  -\frac{e\lambda_V}{m_{\eta}}\frac{g_{VNN}^v}{2m}
\frac{1}{t-m_V^2}\,. 
\end{eqnarray}

(iii) $S_{11}(1535)$ resonance
\begin{eqnarray}
A_1&=&  e\frac{g_{\eta NS_{11}}g_{\gamma\eta S_{11}}}
{m+M_{S_{11}}}(m+M_{S_{11}})\Big(\frac{1}
{s-M_{S_{11}}^2+i\Gamma_{S_{11}} 
M_{S_{11}}}+\frac{1}{u-M_{S_{11}}^2}\Big)\,, \\
A_2&=&0\,,\\
A_3&=&  e\frac{g_{\eta NS_{11}}g_{\gamma\eta S_{11}}}
{m+M_{S_{11}}}\Big(\frac{1}
{s-M_{S_{11}}^2+i\Gamma_{S_{11}} 
M_{S_{11}}}-\frac{1}{u-M_{S_{11}}^2}\Big)\,, \\
A_4&=&  e\frac{g_{\eta NS_{11}}g_{\gamma\eta S_{11}}}
{m+M_{S_{11}}}\Big(\frac{1}
{s-M_{S_{11}}^2+i\Gamma_{S_{11}} 
M_{S_{11}}}+\frac{1}{u-M_{S_{11}}^2}\Big)\,. 
\end{eqnarray}

(iv) $D_{13}(1520)$ resonance
\begin{eqnarray}
A_1&=&  \frac{e g_{\eta ND_{13}}}{s-M_{D_{13}}^2+
i\Gamma M_{D_{13}}}\bigg[\frac{g_{\gamma ND_{13}}^{(1)}}{12mm_{\eta}}
\bigg(3t-m_{\eta}^2-\frac{(\sqrt{s}-m)^2}{s}
(m^2-m_{\eta}^2+m\sqrt{s})\bigg)
\nonumber  \\
& &+\frac{g_{\gamma ND_{13}}^{(2)}}{48m^2m_{\eta}}
\bigg(2m(3t-2m_{\eta}^2)-\frac{(s-m^2)^2}{\sqrt{s}}+\frac{s+m^2}{\sqrt{s}}
m_{\eta}^2\bigg)\bigg]\,, \\
A_2&=&  -\frac{e g_{\eta ND_{13}}}{s-M_{D_{13}}^2
+i\Gamma M_{D_{13}}}
\bigg[\frac{g_{\gamma ND_{13}}^{(1)}}{2m m_{\eta}}
+\frac{g_{\gamma ND_{13}}^{(2)}}{8m^2 m_{\eta}}
(\sqrt{s}+m)\bigg]\,, \\
A_3&=&  \frac{e g_{\eta ND_{13}}}{s-M_{D_{13}}^2+
i\Gamma M_{D_{13}}}
\bigg[\frac{g_{\gamma ND_{13}}^{(1)}}{12m m_{\eta}}
(\sqrt{s}-m)\bigg(2+\frac{(\sqrt{s}+m)^2}{s}+m\frac{m_{\eta}^2}
{s}\bigg)
\nonumber \\ 
& &+\frac{g_{\gamma ND_{13}}^{(2)}}{48m^2 m_{\eta}}
\bigg(3t-2m_{\eta}^2 +(s-m^2)\frac{m+5\sqrt{s}}{\sqrt{s}}+m\frac{m_{\eta}^2}
{\sqrt{s}}\bigg)\bigg]\,, \\ 
A_4&=&  \frac{e g_{\eta ND_{13}}}{s-M_{D_{13}}^2+
i\Gamma M_{D_{13}}}
\bigg[\frac{g_{\gamma ND_{13}}^{(1)}}{12m m_{\eta}}
(\sqrt{s}-m)\bigg(-4+\frac{(\sqrt{s}+m)^2}
{s}+m\frac{m_{\eta}^2}{s}\bigg)
\nonumber \\ 
& &+\frac{g_{\gamma ND_{13}}^{(2)}}{48m^2 m_{\eta}}
\bigg(3t-2m_{\eta}^2+(s-m^2)\frac{m-\sqrt{s}}{\sqrt{s}}+m\frac{m_{\eta}^2}
{\sqrt{s}}\bigg)
\bigg] \,.
\end{eqnarray}

\begin{table}
\caption{
Electromagnetic and hadronic vertices 
of the elementary \protect{$\gamma N \rightarrow \eta N$} model.
The symbols $e_N$ and $\kappa_N$ denote the charge and the anomalous
magnetic moment of the nucleon, respectively ($e_p=1$, $e_n=0$,
and $\kappa_p=1.79$, $\kappa_n=-1.91$). For the $\eta NN$ 
and $\eta NN^{*}$ vertices only ps coupling is considered.
}
\begin{tabular}{cl}
$\gamma NN$
& $  -ee_N\gamma\cdot\varepsilon +i\frac{e\kappa_N}{2m}
\sigma_{\mu\nu}k^{\mu}\varepsilon^{\nu}$ \\
$\eta NN$   & $  -ig_{\eta NN}\gamma_5$ \\
$V\eta\gamma$ & $  \frac{e\lambda}{m_{\eta}}
\varepsilon_{\mu\nu\lambda\sigma}
k^{\nu}\varepsilon^{\lambda}(p^{\prime}-p)^{\sigma}$ \\
$VNN$ & $  -g_{VNN}^v\gamma_{\mu}-i\frac{g_{VNN}^t}{2m} 
\sigma_{\mu\nu}(p^{\prime}-p)^{\nu}$ \\
$\gamma NS_{11}$ & $  ie\frac{g_{\gamma\eta S_{11}}}
{M_{S_{11}}+m}\gamma_5\sigma_{\mu\nu}k^{\mu}\varepsilon^{\nu}$ \\
$\eta NS_{11}$ & $  -ig_{\eta NS_{11}}$ \\
$\gamma ND_{13}$ & $  eg_{\gamma ND_{13}}^{(1)}
\frac{k_{\mu}\gamma\cdot\varepsilon
-\varepsilon_{\mu}\gamma\cdot k}{2m}
+eg_{\gamma ND_{13}}^{(2)}\frac{\varepsilon_{\mu}
k\cdot p-k_{\mu}\varepsilon\cdot p}{4m^2}$ \\
$\eta ND_{13}$ & $  
-i\frac{g_{\eta ND_{13}}}{m_{\eta}}\gamma_5q_{\mu}$\\ 
\end{tabular}
\label{tab1}
\end{table}

\begin{table}
\caption{
Propagators used in the present calculation.
The symbols $M$, $\Gamma$ and $p_\mu$ denote
mass, width and four-momentum of the particle.
}
\begin{tabular}{ll}
spin 1/2 &
$ \frac{\gamma\cdot p+M}{p^2-M^2+iM\Gamma}$ \\
spin 1 & $ \frac{1}{p^2-M^2}\Big[-g_{\mu\nu}+
\frac{p_{\mu}p_{\nu}}{M^2}\Big]$ \\ 
spin 3/2 & $ \frac{\gamma\cdot p+\sqrt{s}}{p^2-M^2+iM\Gamma} 
\Big[g_{\mu\nu}-\frac{\gamma_{\mu}\gamma_{\nu}}{3}-\frac{2}{3}
\frac{p_{\mu}p_{\nu}}{s}-\frac{\gamma_{\mu}p_{\nu}-\gamma_{\nu}
p_{\mu}}{3\sqrt{s}}\Big]$ \\
\end{tabular}
\label{tab2}
\end{table}

\begin{table}
\caption{
Coupling constants of the vector mesons. 
}
\begin{tabular}{lcccc}
Vector meson & Mass [MeV] & $(g_{VNN}^v)^2/4\pi$ & $(g_{VNN}^t)^2/4\pi$ &
$\lambda_V$ \\ \hline
$\rho$  & 770 & 0.55$\pm 0.06$ & $20.5 \pm 2.1$ & $1.06 \pm 0.15$   \\ 
$\omega$  & 782 &$8.1 \pm 1.5$&$0.2 \pm 0.5$&$0.31\pm 0.06$\\ 
\end{tabular}
\label{tab3}
\end{table}

\begin{table}
\caption{
Parameters of the baryon resonances considered in this work.
The values for the $D_{13}(1520)$ resonance are taken from the PDG 1996 
listings \protect{\cite{PDG96}}, 
while those for the $S_{11}(1535)$ are obtained by fitting the experimental 
total cross section data \protect{\cite{Kru95}}. The helicity couplings 
$A^N_{\lambda}$ are given in units of $10^{-3}$GeV$^{-1/2}$. 
}
\begin{tabular}{lcc}
 &    $S_{11}(1535)$    &    $D_{13}(1520)$    \\ \hline
Mass [MeV]             & 1535           & $1515-1530\,(1520)$ \\
$\Gamma$ [MeV]         & 160            & $110-135\,(120)$    \\
$\Gamma_{\eta N}$ [MeV]& 80             &  0.12               \\
$A^p_{1/2}$            & 104            & $-24 \pm 9$         \\
$A^n_{1/2}$            & $-88$           & $-59 \pm 9$         \\
$A^p_{3/2}$            &                & $166 \pm 5$          \\
$A^n_{3/2}$            &                & $-139 \pm 11$       \\ 
\end{tabular}
\label{tab4}
\end{table}

\begin{table}
\caption{
Parameters of the $^4$He wave function from \protect{\cite{Sher83}}. 
}
\begin{tabular}{c|ccccc}
   $j$   &   1   &   2   &   3   &   4   &   5   \\ 
\hline
$a_j$ [fm$^{-1}$]& 1.305 & $-5.222$ & 7.832 & $-5.222$ & 1.305    \\ 
$\beta_j$ [fm$^{-1}$] & 0.0 & 1.42 & 2.84 & 4.26 & 5.68 \\ 
\end{tabular}
\label{tab5}
\end{table}

\begin{figure}
\centerline{\psfig{figure=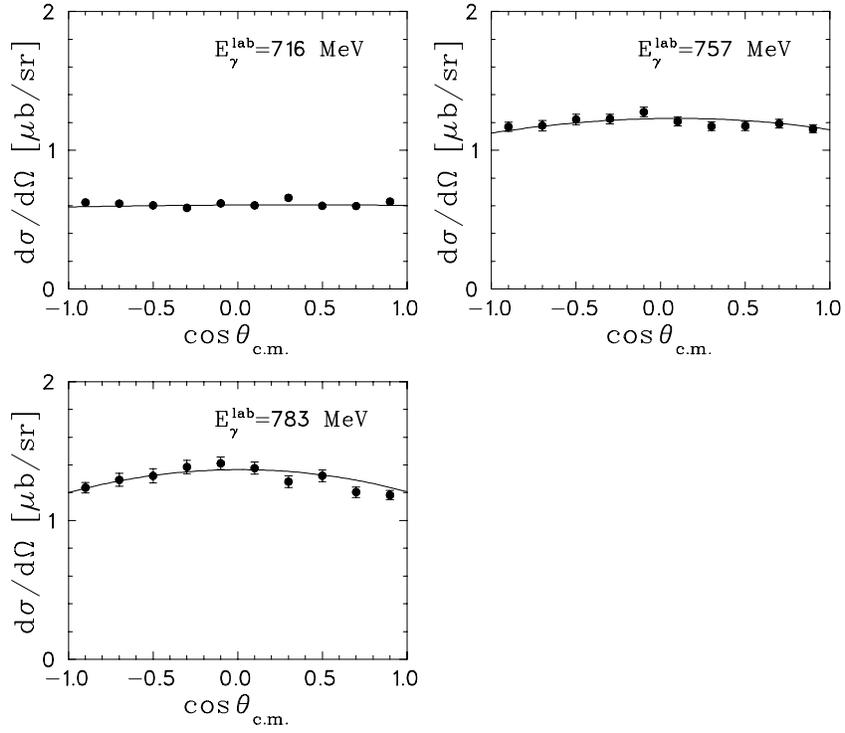,width=11cm,angle=0}}
\vspace*{.5cm}
\caption{Comparison of the elementary model for $\gamma p \rightarrow \eta 
p$ with the Mainz data \protect{\cite{Kru95}}. 
The curves were obtained by fitting the $S_{11}(1535)$ parameters.}
\label{fig2}
\end{figure}

\begin{figure}
\centerline{\psfig{figure=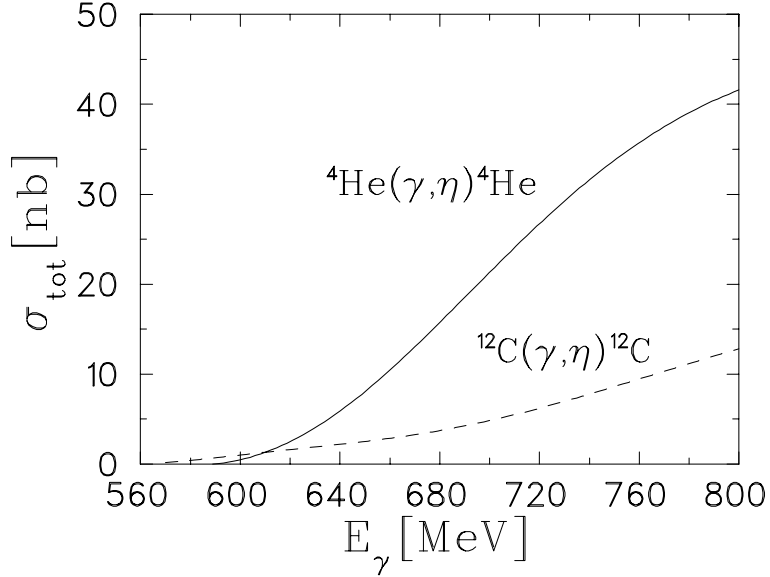,width=10cm,angle=0}}
\vspace*{.5cm}
\caption{Total cross section for coherent $\eta$-photoproduction
on $^4$He and $^{12}$C.}
\label{figTot}
\end{figure}

\begin{figure}
\centerline{\psfig{figure=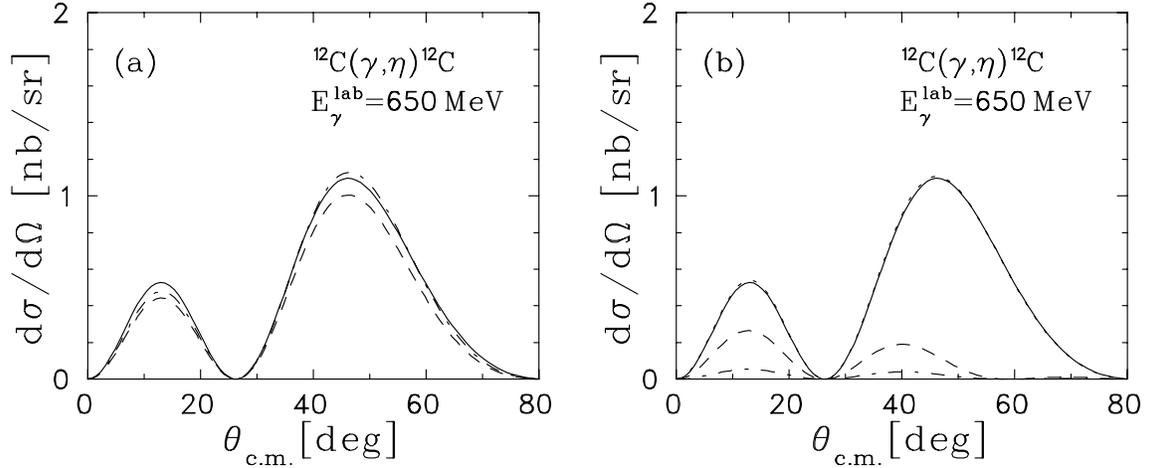,width=15cm,angle=0}}
\vspace*{.5cm}
\caption{Differential cross section for $^{12}$C$(\gamma,\eta)^{12}$C. 
The curves show the contributions from different terms included in  
the elementary $\gamma N \rightarrow \eta N$ model: (a) 
full calculation (solid), $\omega$ exchange only (dashed),
without nucleon pole terms (dash-dotted); (b) 
full calculation (solid), without $S_{11}(1535)$ (dotted), only 
$D_{13}(1520)$ with free width $\Gamma$ (dashed) and with medium 
modified width $\Gamma^*$ (dash-dotted).}
\label{C12_1}
\end{figure}

\begin{figure}
\centerline{\psfig{figure=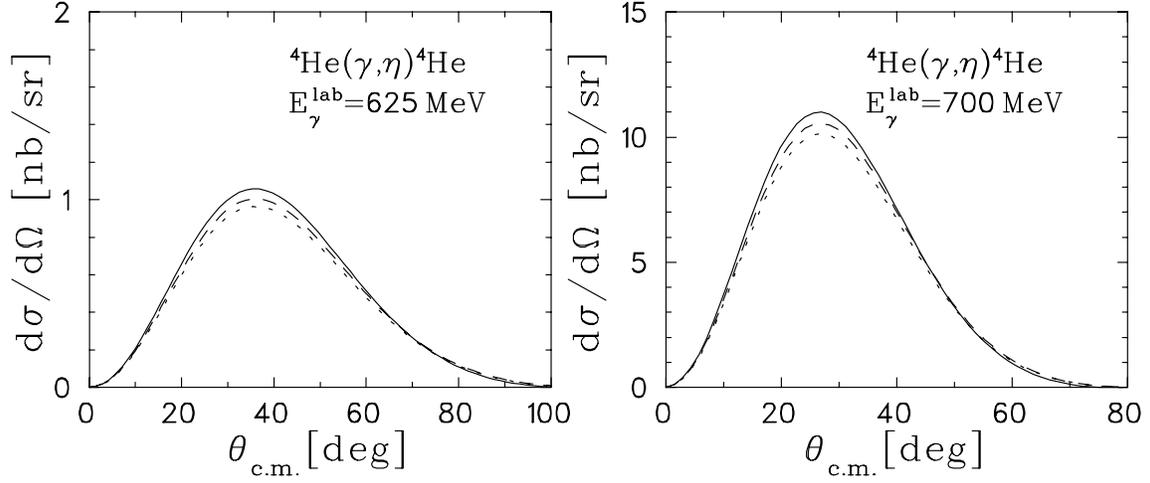,width=15cm,angle=0}}
\vspace*{.5cm}
\caption{
Differential cross section for the reaction $^4$He$(\gamma,\eta)^4$He 
calculated with inclusion of Fermi motion and by taking as invariant 
energy of the $\gamma N$ subsystem $W_{\gamma N}^{(1)}$ (dashed curves)
and $W_{\gamma N}^{(2)}$ (dotted curves). The solid curves are the result of 
the factorization approximation (\protect{\ref{8f}}) with on-shell 
amplitude.
}
\label{W}
\end{figure}

\begin{figure}
\centerline{\psfig{figure=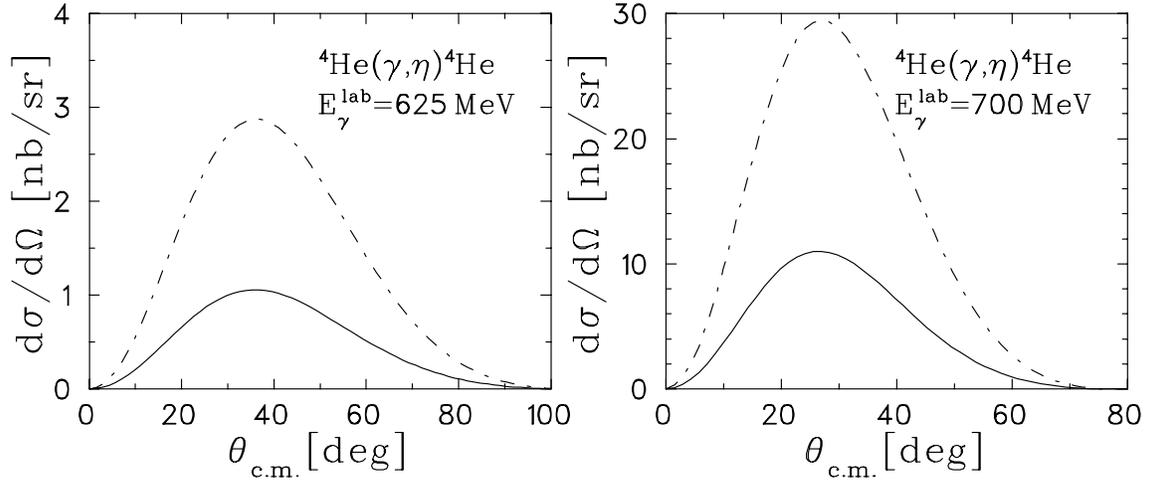,width=15cm,angle=0}}
\vspace*{.5cm}
\caption{
Influence of the $\omega$-nucleon coupling constant on the differential cross 
section for $^4$He$(\gamma,\eta)^4$He. Full curve  shows the cross section 
calculated with $g_{\omega NN}^v=10.3$, and the dash-dotted curve for 
$g_{\omega NN}^v=17.5$ from Ref.\ \protect{\cite{Laget77}}.}
\label{W1}
\end{figure}

\begin{figure}
\centerline{\psfig{figure=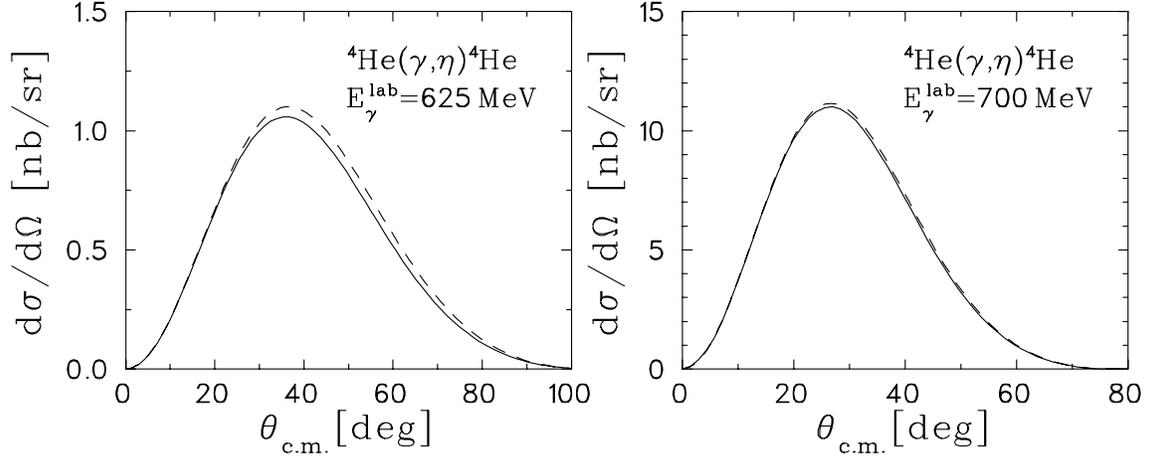,width=15cm,angle=0}}
\vspace*{.5cm}
\caption{ Comparison of $^4$He$(\gamma,\eta)^4$He cross sections calculated 
in the c.m.\ frame (full) and in the lab frame (dashed) for two photon 
energies.}
\label{figlabcm}
\end{figure}

\begin{figure}
\centerline{\psfig{figure=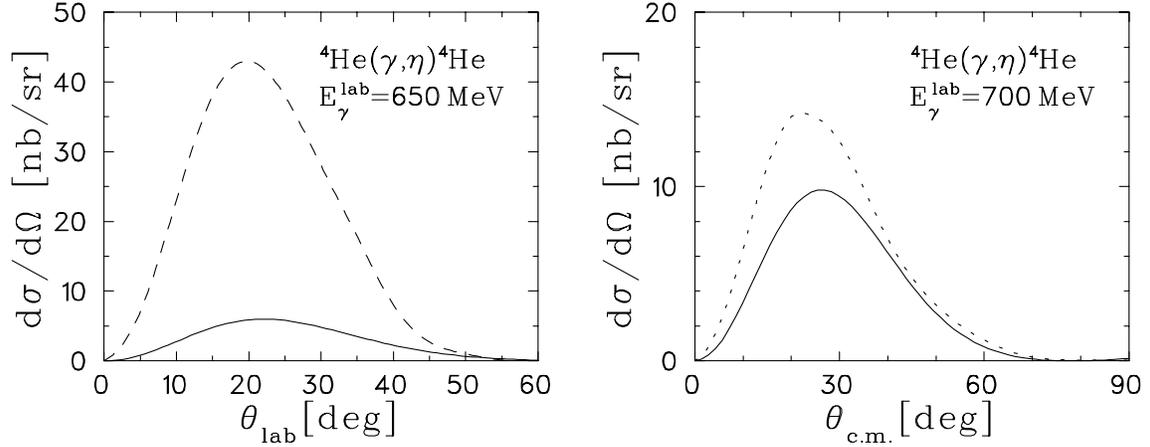,width=15cm,angle=0}}
\vspace*{.5cm}
\caption{Comparison of the angular distributions for $^4$He$(\gamma,\eta)^4$He 
obtained in the present work (full curves) with the calculations of 
Ref.\ \protect{\cite{Bennh91}} (left: dashed) and 
Ref.\ \protect{\cite{Tiat94}} (right: dotted). 
}
\label{Bennh}
\end{figure}

\begin{figure}
\centerline{\psfig{figure=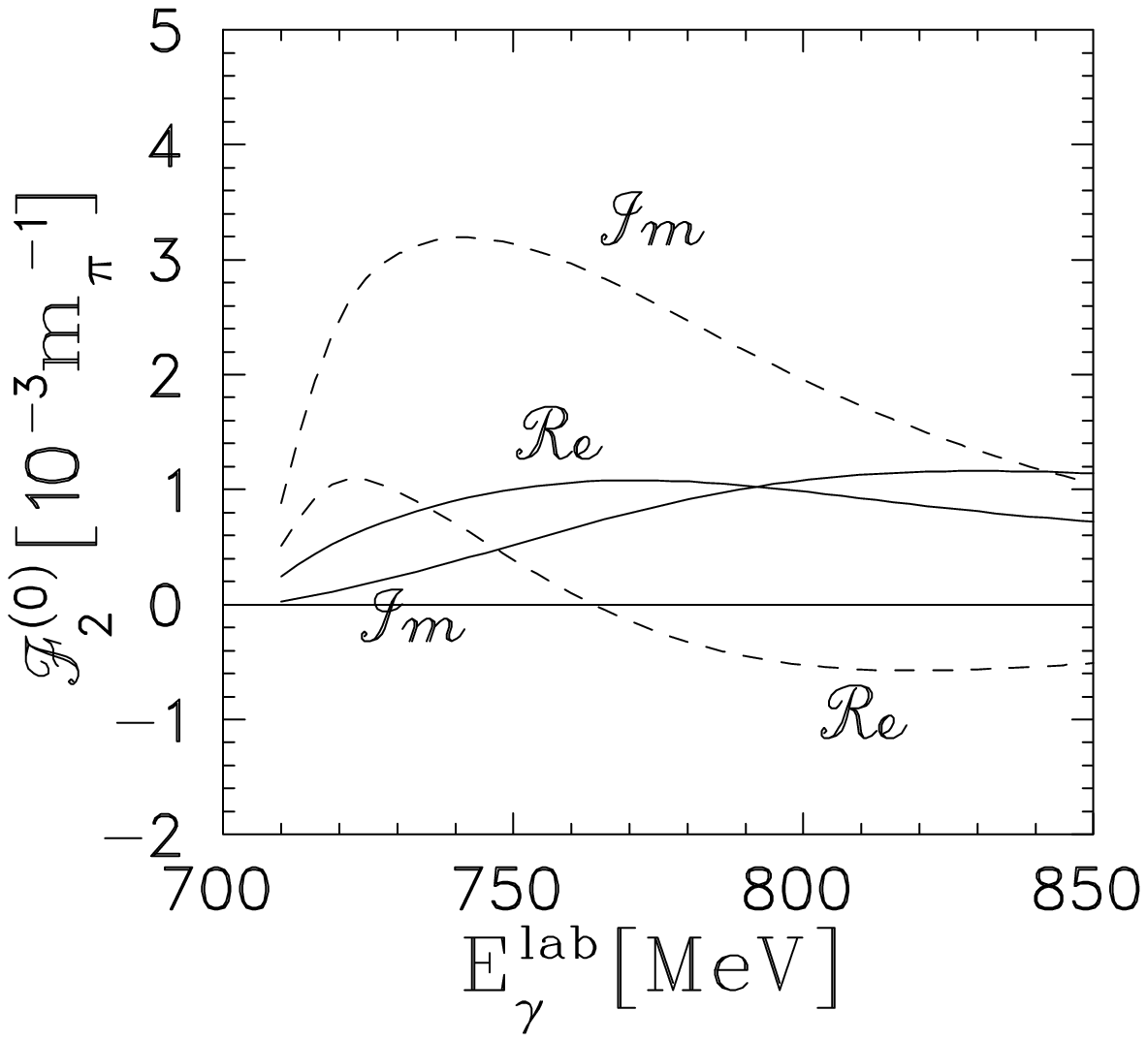,width=10cm,angle=0}}
\vspace*{.5cm}
\caption{
Comparison of the real and imaginary parts of the CGLN isoscalar amplitude 
${\cal F}_2^{(0)}(W,\theta)$ for $\theta=30^\circ$ as predicted in 
Ref.\ \protect{\cite{Bennh91}} (dashed) and 
by our model (solid).}
\label{Bennh2}
\end{figure}


\begin{thebibliography}{99}

\bibitem{Kru95}
B.\ Krusche et al., Phys.\ Rev.\ Lett.\ {\bf 74} (1995) 3736

\bibitem{Mukh95}
M.\ Benmerrouche, N.C.\ Mukhopadhyay, J.F.\ Zhang, Phys.\ Rev.\ D {\bf 51}
(1995)  3237  

\bibitem{Sau95}
Ch.\ Sauermann, B.L.\ Friman, W.\ N\"orenberg, Phys.\ Lett.\ {\bf B341}
(1995) 261; GSI-Preprint-97-03, nucl-th/9701022

\bibitem{Breit97}
E.\ Breitmoser, H.\ Arenh\"ovel, Nucl.\ Phys.\ {\bf A612} (1997) 321
\bibitem{Car93}
R.C.\ Carrasco, Phys.\ Rev.\ C {\bf 48} (1993) 2333

\bibitem{Str}
TAPS-Experiment at MAMI (1996) 
(Proposal A2/12-93, Spokesperson: B.\ Krusche)

\bibitem{Bennh91}
C.\ Bennhold, H.\ Tanabe, Nucl.\ Phys.\ {\bf A530} (1991) 625

\bibitem{Bennh89}
C.\ Bennhold, Phys.\ Rev.\ C {\bf 39} (1989)  1944  

\bibitem{Bjor64}
J.D.\ Bjorken, S.D.\ Drell, Relativistic Quantum Mechanics 
(McGraw-Hill, New York, 1964) 

\bibitem{CGLN}
G.F.\ Chew, M.L.\ Goldberger, F.E.\ Low, Y.\ Nambu,  Phys.\ Rev.\ {\bf 106}
(1957) 1345

\bibitem{Tiat94}
L.\ Tiator, C.\ Bennhold, S.S.\ Kamalov, Nucl.\ Phys.\ {\bf A580} (1994) 455 

\bibitem{KiT96}
M.\ Kirchbach, L.\ Tiator, Nucl.\ Phys.\ {\bf A604} (1996) 385 

\bibitem{Dum83}
O.\ Dumbrajs, R.\ Koch, H.\ Pilkuhn, G.C.\ Oades, H.\ Behrens, J.J.\ De 
Swart, P.\ Kroll, Nucl.\ Phys.\ {\bf B216} (1983)  277 

\bibitem{Dol}
S.I.\ Dolinsky et al., Z. \ Phys.\ C {\bf 42} (1989) 511

\bibitem{PDG96}
Review of Particle Properties, Phys.\ Rev.\ D {\bf 54} (1996) 1

\bibitem{Fix97}
A.I.\ Fix, V.A.\ Tryasuchev, Yad.\ Fiz.\ {\bf 60} (1997) 41 
(Phys.\ Atom.\ Nucl.\ {\bf 60} (1997) 35)

\bibitem{Will85}
H.T.\ Williams, Phys.\ Rev.\ C {\bf 31} 2297  (1985)

\bibitem{Adel85}
R.A.\ Adelseck, C.\ Bennhold, L.E.\ Wright,  Phys.\ Rev.\ C {\bf 32} 
1681  (1985) 

\bibitem{Pol73}
Yu.N.\ Eldyshev, V.N.\ Lukyanov, Yu.S.\ Pol, Yad.\ Fiz.\ {\bf 16} (1972) 
506 (Sov.\ J.\ Nucl.\ Phys.\ {\bf 16} (1973) 282)

\bibitem{Kam87}
A.A.\ Chumbalov, R.A.\ Eramzhyan, S.S.\ Kamalov, Z.\ Phys.\ {\bf A328} 
(1987) 195  

\bibitem{Sher83}
H.S.\ Sherif, M.S.\ Abdelmonem, R.S.\ Sloboda,  Phys.\ Rev.\ C {\bf 27} 
(1983) 2759    

\bibitem{Bian93}
N.\ Bianchi et al., Phys.\ Lett.\ {\bf B299} (1993) 219 \\
M.\ Anghinolfi et al., Phys.\ Rev.\ C {\bf 47} (1993) R922

\bibitem{KruPR95}
B.\ Krusche, Proceedings of Workshop on Hadrons in Nuclear Matter, 
Hirschegg, Austria, eds.\ H.\ Feldmeier and W.\ N\"orenberg (GSI,
Darmstadt 1995)

\bibitem{Alb94}
W.M.\ Alberico, G.\ Gervino, A.\ Lavagno, Phys.\ Lett.\ {\bf B321} 
(1994) 177

\bibitem{PiS97}
J.\ Piekarewicz, A.J.\ Sarty, M.\ Benmerrouche, preprint, nucl-th/9701019

\bibitem{Tiat80}
L.\ Tiator, A.K.\ Rej, D.\ Drechsel, Nucl.\ Phys.\ {\bf A333} (1980) 343

\bibitem{PDG}
P.\ S\"oding et al.,  Phys.\ Lett.\  {\bf B39} (1972) 1

\bibitem{Laget77}
I.\ Blomqvist, J.M.\ Laget,  Nucl.\ Phys.\ {\bf A280} (1977) 405

\end{thebibliography}
\end{document}